\documentstyle[psfig,conf_iap,]{article}

\begin{document}

\heading{Interstellar Abundances in the Magellanic Clouds}

\par\medskip\noindent

\author{
D. E. Welty$^{1}$
}
\address{University of Chicago, AAC, 5640 S. Ellis Ave., Chicago, IL 60637}

\begin{abstract}
We summarize recent studies of the predominantly neutral gas in the Magellanic Clouds (MC), based on combinations of high-resolution optical spectra and UV spectra obtained with {\it IUE}, {\it HST}, and {\it FUSE}.
Similarities in relative gas-phase abundances for Galactic and MC sightlines with relatively low $N$(H) had suggested that the depletion patterns might be similar in the three galaxies --- at least where the depletions are relatively mild.
New STIS spectra of one higher $N$(H) SMC sightline, however, have revealed significant differences, relative to the Galactic patterns, for SMC clouds characterized by more severe depletions.
\end{abstract}

\section{Introduction}

Studies of the interstellar medium in the Magellanic Clouds (MC) explore different environmental conditions from those typically probed in our own Galactic ISM. 
On average, MC stars and H~II regions have metallicities below solar values by 0.3 dex (LMC) and 0.6--0.7 dex (SMC), though the {\it relative} elemental abundances (e.g., [X/Zn], for elements X that we consider) are generally similar to those found for analogous Galactic objects (\cite{rd92}; \cite{wlb97}; \cite{wfs99}; \cite{ven99}).
The ISM in both MC generally has lower dust-to-gas ratios, stronger ambient radiation fields, and significant differences in UV extinction (especially in the SMC).
Determination of elemental abundances, depletions, and physical conditions in the MC ISM therefore provides interesting tests for theoretical models of interstellar clouds and dust grains, and should also yield insights into the nature of the even lower metallicity QSO absorption-line systems.

Most studies of the optical absorption lines of Ca~II and Na~I have focused on the kinematics and structure of the MC (e.g., \cite{way90}; but see \cite{vmm93}).
While UV spectra of many MC stars were obtained with {\it IUE}, the relatively low resolution (FWHM $\sim$ 20--25 km~s$^{-1}$) and generally low S/N made it very difficult to determine accurate abundances and physical conditions for individual interstellar clouds in the generally complex MC sightlines (e.g., \cite{fs83}; \cite{dfs85}). 

New observing capabilities (both on the ground and in space) are enabling much more detailed studies of the MC ISM, however (e.g., \cite{iau190}).
In this contribution, we focus on recent studies of the predominantly neutral gas in the MC, based on high-resolution optical spectra and on UV spectra from {\it IUE}, {\it HST} (GHRS and STIS), and {\it FUSE}.
We compare relative gas-phase abundances [X/Zn] (which essentially give the depletion of X) in the MC ISM with representative patterns found in our Galaxy (\cite{ss96}; \cite{wlb97}) and note constraints on the physical conditions that can be derived from the available data.

\begin{figure}
\centerline{\vbox{
\psfig{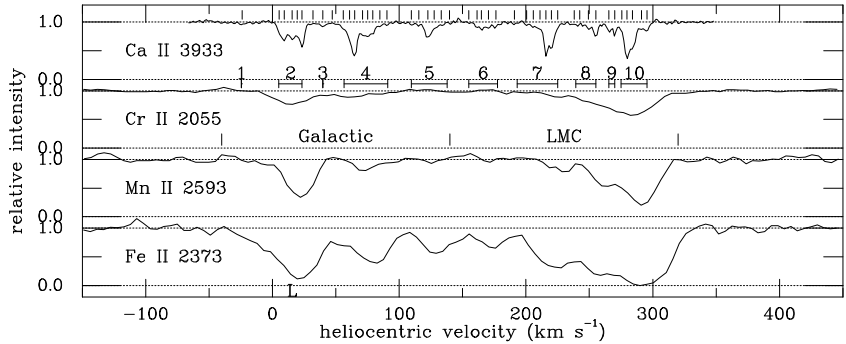}
}}
\caption[]{Optical and UV spectra of SN~1987A (\cite{vac87}; \cite{wfs99}).
Component groups 1--5, at $-$25 km~s$^{-1}$ $\le$ $v$ $\le$ 140 km~s$^{-1}$, are due to gas in the Galactic disk and halo.
Groups 6--10, at 155 km~s$^{-1}$ $\le$ $v$ $\le$ 295 km~s$^{-1}$, are due to LMC gas.}
\end{figure}

\section{Large Magellanic Cloud}

{\bf SN~1987A:}
High-resolution (FWHM $\sim$ 3 km~s$^{-1}$) Na~I and Ca~II spectra (\cite{vac87}) reveal at least 46 components, at velocities between $-$25 and +295 km s$^{-1}$, which may be associated with 10 component groups discernible in the lower resolution {\it IUE} spectra (\cite{wfs99}) (Fig.~1).
Accurate component group column densities for many neutral and singly ionized species were obtained by using the detailed component structure found for Na~I and Ca~II to model the UV line profiles of similarly distributed species.
The main LMC component groups (8--10), with total $N$(H) $\sim$ 2.8 $\times$ 10$^{21}$ cm$^{-2}$, have relative abundances [X/Zn] similar to those found for Galactic ``warm, diffuse'' clouds (Fig.~2).
Variations in individual component $N$(Na~I)/$N$(Ca~II), narrow line widths for individual components seen in even higher resolution Na~I and K~I spectra (FWHM $\sim$ 0.5 km~s$^{-1}$; \cite{pg88}), and analysis of the C~I fine structure excitation equilibrium, however, strongly suggest that the main LMC groups contain both warm and cold gas. 
More severe depletions in the colder gas could be masked by blending with components with less severe depletions.
The lower column density LMC groups (6--7), with $N$(H) $\sim$ 1.4 $\times$ 10$^{20}$ cm$^{-2}$, have [X/Zn] more similar to the values found for clouds in the Galactic halo.

{\bf [Cr/Zn]:} 
Column densities derived from {\it HST}/GHRS echelle spectra (\cite{rb97}) yield [Cr/Zn] = $-$0.76 for the LMC components toward Sk~$-$69$^o$175 --- suggesting depletions similar (on average) to those for group 10 toward SN~1987A.
More severe depletions are likely toward Sk~$-$68$^o$73, which has [Cr/Zn] = $-$1.13, as well as stronger Na~I, Mg~I, and Zn~II.

{\bf Sk~$-$67$^o$5:} 
Spectra from {\it FUSE} (\cite{fri00}) and {\it IUE} yield relative abundances [X/Zn] (for X = Si, P, Cr, Fe, and Ni) similar to those for group 10 toward SN~1987A.
The {\it FUSE} spectra also show strong absorption from H$_2$, with $N$(H$_2$) $\sim$ 3 $\times$ 10$^{19}$ cm$^{-2}$, $T_{01}$ $\sim$ 60 K, and $T_{\rm ex}$ $\sim$ 800 K for $J$ = 3--5 (\cite{fri00}).

\begin{figure}
\centerline{\vbox{
\psfig{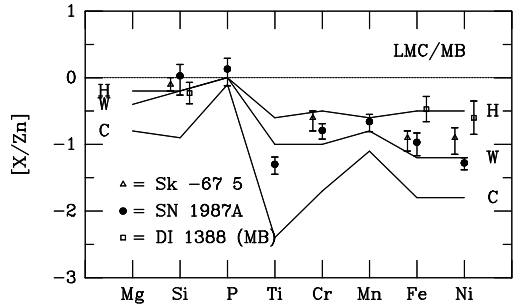}
}}
\caption[]{Relative abundances found for main LMC components toward SN 1987A and Sk$-$67$^o$5 (\cite{wfs99}; \cite{fri00}), and for main Magellanic Bridge components toward DI 1388 (\cite{lsd01}; we assume [S/Zn] = 0.1).
Solid lines show representative patterns for Galactic ``cold, dense'' clouds (C), ``warm, diffuse'' clouds (W), and halo clouds (H) (\cite{ss96}; \cite{wlb97}).}
\end{figure}

\section{Magellanic Bridge}

STIS echelle spectra (FWHM $\sim$ 6.5 km~s$^{-1}$) have been obtained for DI 1388, located in the Magellanic Bridge, where the overall metallicity has been estimated as about 1.1 dex below solar (\cite{lsd01}).
For two components with total $N$(H) $\sim$ 1--2 $\times$ 10$^{20}$ cm$^{-2}$, the relative abundances [X/S] (for X = Al, Si, Fe, and Ni) are similar to those found for Galactic halo clouds (Fig.~2).
The apparent underabundance of N~I in the weaker component may be due to partial ionization.
Combining a (tentative) detection of C~I with a limit on $n_e$ (from analysis of the C~II fine structure excitation) suggests either a relatively weak radiation field or a relatively low $T$ for the stronger component.

\begin{figure}
\centerline{\vbox{
\psfig{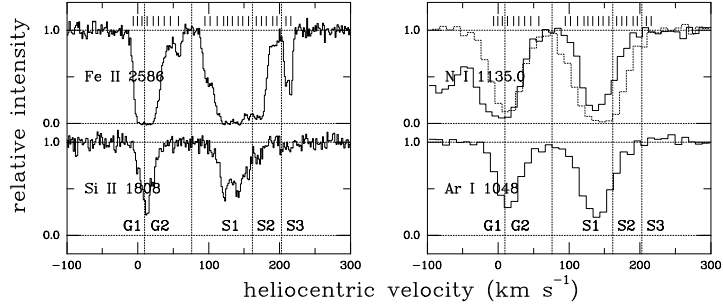}
}}
\caption[]{GHRS ({\it left}) and {\it FUSE} ({\it right}) spectra of SMC star Sk~108 (\cite{wlb97}; \cite{mal01}).
Component groups G1 and G2 are due to gas in Galactic halo and disk; groups S1, S2, and S3 are due to SMC gas.
Dotted line shown with N~I $\lambda$1135 is the profile for O~I $\lambda$1039, which is stronger (relatively) than would be expected for solar N/O.}
\end{figure}

\section{Small Magellanic Cloud}

{\bf Sk~108:}
GHRS echelle spectra (FWHM $\sim$ 4.2 km~s$^{-1}$) reveal at least 16 SMC components between 89 and 210 km~s$^{-1}$ (Fig.~3; \cite{wlb97}).
For all the SMC components, the relative abundances [X/Zn] (for X = Si, Cr, Mn, Fe, and Ni) are broadly similar to those in Galactic halo clouds, though Si may be slightly high and Ni slightly low (Fig.~5).
For $N$(H) $\sim$ 3.5 $\times$ 10$^{20}$ cm$^{-2}$, the overall [Zn/H] is consistent with a metallicity 0.6 dex below solar and a Zn depletion of 0.1 dex (as in the halo clouds).
Analysis of lower resolution GHRS G160M and {\it FUSE} spectra (FWHM $\sim$ 20--25 km~s$^{-1}$) suggests that P, S, and Ar are essentially solar, relative to Zn (\cite{mal01}).
The abundance of N may be low, but is uncertain due to saturation of the N~I lines near 1135\AA, possible partial ionization in the outlying SMC components, and uncertainties in the overall SMC N abundance.
Strong absorption due to N~III, S~III, and Fe~III indicates a significant ionized component in the SMC.
Weak absorption from the $J$ = 1 and $J$ = 3 levels of H$_2$ appears consistent with $N_{\rm tot}$(H$_2$) $\sim$ 3 $\times$ 10$^{14}$ cm$^{-2}$ and $T_{\rm ex}$ $\sim$ 1000 K (for all levels); Na~I and C~I are not detected.

{\bf Sk~78 (HD~5980):}
STIS echelle spectra (FWHM $\sim$ 6.5 km~s$^{-1}$) were obtained to study the post-outburst spectrum of this WR binary (\cite{koe00}); many {\it IUE} spectra are also available (\cite{fs83}).
The overall SMC relative abundances [X/Zn] (for X = S, Ti, Cr, Mn, Fe, and Ni) fall between the ``warm, diffuse'' and halo cloud patterns; Mg and Si are slightly higher (Fig.~5; \cite{wlb01}).
{\it FUSE} spectra indicate $N$(H$_2$) $\sim$ 4.2 $\times$ 10$^{15}$ cm$^{-2}$, compared to a total SMC $N$(H~I) $\sim$ 8.5 $\times$ 10$^{20}$ cm$^{-2}$ (\cite{shu00}); Na~I and C~I are detected, but weak.

{\bf Sk~155:}
STIS echelle spectra reveal at least 12 Milky Way and 25 SMC components, over a total velocity range of nearly 300 km~s$^{-1}$ (Fig.~4; \cite{wlb01}); $N$(H) $\sim$ 3.5 $\times$ 10$^{21}$ cm$^{-2}$.
Striking differences between the profiles of trace neutral species, dominant depleted species (e.g., Fe~II, Ni~II), and dominant undepleted species (e.g., S~II, Zn~II) imply significant variations in depletions and physical conditions among the three main SMC component groups (A, B, C).
The depletions of Fe and Ni range from mild ([Fe,Ni/Zn] $\sim$ $-$0.3 to $-$0.8 dex) to severe ([Fe,Ni/Zn] $\sim$ $-$1.7 dex); Mg and Si, however, are essentially undepleted ([Mg,Si/Zn] $\sim$ $-$0.1 to +0.2 dex) throughout.
The combination of severe depletion of Cr, Mn, Fe, and Ni and negligible depletion of Mg and Si has not been seen in any Galactic sightline --- but is consistent with the [Mg,Si/Zn] found toward Sk~108 and Sk~78 (Fig.~5).
Both C~I and its excited fine structure levels are detected; analysis of the excitation equilibrium suggests pressures $n_{\rm H}T$ $\sim$ 1--2 $\times$ 10$^4$ cm$^{-3}$K --- significantly higher than values typically found in the Galactic ISM (\cite{jt01}), as predicted by theoretical models for clouds in low metallicity, high radiation environments (\cite{whm95}).
Intriguingly, the strongest Cl~I (which should track H$_2$) is not in the component with the most severe depletions of Fe and Ni --- which may suggest differences in the relationship between molecular fraction and depletions in the SMC.

\begin{figure}
\centerline{\vbox{
\psfig{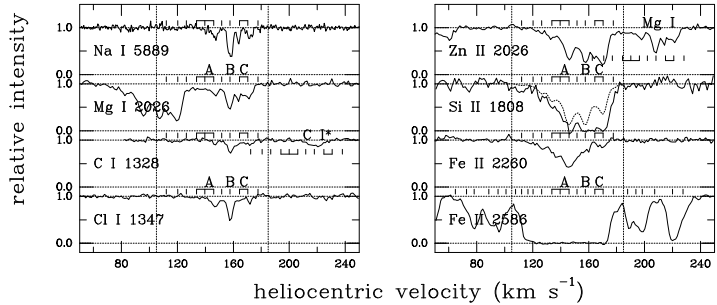}
}}
\caption[]{Optical and UV spectra of SMC components toward Sk~155 (\cite{wlb01}).
Note dramatic differences in the profiles for trace neutral species, Fe~II (dominant depleted), and Zn~II and Si~II (dominant undepleted).
Dotted line for Si~II is the profile predicted from Fe/Zn if Galactic depletions are assumed.}
\end{figure}

\section{Discussion/Summary}

The relative abundances in the LMC gas (total line of sight values toward four stars) are broadly similar to those found for warm, diffuse clouds in the Galactic disk and halo.
The available diagnostics (e.g., higher resolution Na~I and Ca~II spectra, C~I fine structure) suggest that the individual components are likely characterized by wider ranges in depletions and physical conditions, however.
Higher resolution UV spectra for three SMC sightlines do reveal significant differences in depletions (from mild to severe) and physical conditions for individual components --- both within the SMC and relative to the Galactic ISM.
We conclude with some implications and areas for further study:

If Si is generally undepleted in the SMC, then models of SMC dust --- which rely heavily on silicates (\cite{wd01}) --- will need revision; oxides and/or metallic Fe grains may dominate.
If Si is also undepleted in QSO absorption-line systems (which resemble the SMC ISM in some respects), then observed QSOALS [Si/Fe] ratios would not imply enhanced $\alpha$-element/Fe ratios.
The higher pressures found toward Sk 155 are consistent with theoretical models; are such elevated pressures characteristic of the SMC ISM?
General weakness of the trace neutral species (e.g., Na~I, C~I) in the MC, relative to H --- even after accounting for the lower metallicity and stronger radiation fields --- may provide further evidence for additional processes affecting the ionization of heavy elements in the ISM (\cite{wh01}).
A new {\it FUSE} survey of H$_2$ toward 70 MC stars has revealed some differences in the behavior of H$_2$, relative to that in the Galactic ISM (\cite{tum01}); comparisons of depletions and H$_2$, for the different environmental conditions in the Milky Way and MC, should give insight into the relationships between H$_2$, dust, and local physical conditions.

\acknowledgements{We are grateful for support from NASA grants HST-GO-08145.01-A (via STScI) and NAG5-3228 (LTSA), both to the University of Chicago.}

\begin{figure}
\centerline{\vbox{
\psfig{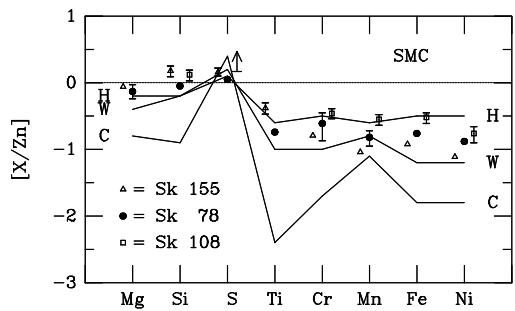}
}}
\caption[]{Relative abundances found for all SMC components toward Sk 108, Sk 78, and Sk 155 (\cite{wlb97}; \cite{mal01}; \cite{wlb01}).
More severe depletions ([Fe,Ni/Zn] $\sim$ $-$1.7) are seen for individual components toward Sk~155.
Note that Mg and Si are higher than the Galactic patterns in all three SMC sightlines.}
\end{figure}

\vfill

\begin{iapbib}{99}{
\bibitem{iau190} New Views of the Magellanic Clouds (IAU Symposium \#190), eds. Y.-H. Chu, N. B. Suntzeff, J. E. Hesser, \& D. A. Bohlender (San Francisco: PASP)
\bibitem{dfs85} de Boer, K. S., Fitzpatrick, E. L., \& Savage, B. D., 1985, MNRAS, 217, 115
\bibitem{fs83} Fitzpatrick, E. L., \& Savage, B. D., 1983, \apj, 267, 93
\bibitem{fri00} Friedman, S. D. et al., 2000, \apj, 538, L39
\bibitem{jt01} Jenkins, E. B., \& Tripp, T. M., 2001, \apj, in press
\bibitem{koe00} Koenigsberger, G. et al., 2000, \aj, 121, 267
\bibitem{lsd01} Lehner, N., Sembach, K. R., Dufton, P. L., Rolleston, W. R. J., \& Keenan, F. P., 2001, \apj, 551, 781
\bibitem{mal01} Mallouris, C. et al., 2001, \apj, 558, 133
\bibitem{pg88} Pettini, M., \& Gillingham, P., 1988, AAO Newsletter \#43
\bibitem{rb97} Roth, K. C., \& Blades, J. C., 1997, \apj, 474, L95
\bibitem{rd92} Russell, S. C., \& Dopita, M. A., 1992, \apj, 384, 508
\bibitem{ss96} Savage, B. D., \& Sembach, K. R., 1996, ARA\&A, 34, 279
\bibitem{shu00} Shull, J. M. et al., 2000, \apj, 538, L73
\bibitem{tum01} Tumlinson, J. et al., 2001, \apj, submitted
\bibitem{ven99} Venn, K. A., 1999, \apj, 518, 405
\bibitem{vac87} Vidal-Madjar, A., Andreani, P., Cristiani, S., Ferlet, R., Lanz, T., \& Vladilo, G., 1987, \aeta, 177, L17
\bibitem{vmm93} Vladilo, G., Molaro, P., Monai, S., D'Odorico, S., Ferlet, R., Vidal-Madjar, A., \& Dennefeld, M., 1993, \aeta, 274, 37
\bibitem{way90} Wayte, S. R., 1990, \apj, 355, 473
\bibitem{wd01} Weingartner, J. C., \& Draine, B. T., 2001, \apj, 548, 296
\bibitem{wfs99} Welty, D. E., Frisch, P. C., Sonneborn, G., \& York, D. G., 1999, \apj, 512, 636
\bibitem{wh01} Welty, D. E., \& Hobbs, L. M., 2001, ApJS, 133, 345
\bibitem{wlb97} Welty, D. E., Lauroesch, J. T., Blades, J. C., Hobbs, L. M., \& York, D. G., 1997, \apj, 489, 672
\bibitem{wlb01} Welty, D. E., Lauroesch, J. T., Blades, J. C., Hobbs, L. M., \& York, D. G., 2001, \apj, 554, L79 (see also these proceedings)
\bibitem{whm95} Wolfire, M. G., Hollenbach, D., McKee, C. F., Tielens, A. G. G. M., \& Bakes, E. L. O., 1995, \apj, 443, 152
}
\end{iapbib}
\end{document}